\documentclass[10pt,english,final, 5p, preprint, times,numbers, sort&compress]{elsarticle}
\usepackage[T1]{fontenc}
\usepackage{array}
\usepackage{amsthm}
\usepackage{amsmath}
\usepackage{amssymb}
\usepackage{graphicx}

\makeatletter

\newcommand{\noun}[1]{\textsc{#1}}
\providecommand{\tabularnewline}{\\}

\numberwithin{equation}{section}
\numberwithin{figure}{section}

\makeatother

\usepackage{babel}
\begin{document}

\title{Three-term Method and Dual Estimate on Static Problems of Continuum
Bodies\tnoteref{tn1}}

\author[a1]{Masaaki Miki\corref{cor1}\fnref{fnd1}}

\ead{mikity@iis.u-tokyo.ac.jp}

\address[a1]{Department of Architecture, School of Engineering, the University
of Tokyo, komaba4-6-1, meguro-ku, Tokyo, 153-8505, JAPAN}

\tnotetext[tn1]{This work supplements Ref. \citep{miki2010}, particularly about
computational algorithms which was used in it.}

\cortext[cor1]{Corresponding author: Tel.: +81 (0)354526403; fax: +81 (0)354526405.}

\fntext[fnd1]{Research Fellow (DC), Japan Society for Promotion of Science}

\begin{abstract}
This work aims to provide standard formulations for direct minimization
approaches on various types of static problems of continuum mechanics.
Particularly, form-finding problems of tension structures are discussed
in the first half and the large deformation problems of continuum
bodies are discussed in the last half. In the first half, as the standards
of iterative direct minimization strategies, two types of simple recursive
methods are presented, namely the two-term method and the three-term
method. The dual estimate is also introduced as a powerful means of
involving equally constraint conditions into minimization problems.
As examples of direct minimization approaches on usual engineering
issues, some form finding problems of tension structures which can
be solved by the presented strategies are illustrated. Additionally,
it is pointed out that while the two-term method sometimes becomes
useless, the three-term method always provides remarkable rate of
global convergence efficiency. Then, to show the potential ability
of the three-term method, in the last part of this work, some principle
of virtual works which usually appear in the continuum mechanics are
approximated and discretized in a common manner, which are suitable
to be solved by the three-term method. Finally, some large deformation
analyses of continuum bodies which can be solved by the three-term
method are presented. \end{abstract}
\maketitle

\section{Introduction}

Within this work, standard formulations for solving various types
of static problems of continuum bodies by the direct minimization
methods are presented. The direct minimization methods are always
associated with static mechanics via \textbf{principle of virtual
work}. For example, the direct minimization approaches are sometimes
very effective on solving form-finding problems of tension structures\citep{miki2010}.
In particular, the aim of this work is to present the basic strategies
such as the \textbf{three-term method} and the \textbf{dual estimate},
and to illustrate various types of static problems that can be solved
by using them.

In section 2, as the standard recursive direct minimization methods,
the \textbf{two-term method} and\textbf{ }the \textbf{three-term method}
are described. While the former is basically identical with the steepest
decent method and the latter is with the dynamic relaxation method,
some differences are pointed out. In addition, via discussion of a
form-finding problem of a simple cable-net structure as a typical
example, the relation over the \textbf{principle of virtual work},
the \textbf{stationary condition}, and the \textbf{standard search
direction} is clarified. Furthermore, the \textbf{dual estimate} is
proposed as a powerful means of involving constraint conditions into
direct minimization approaches. Then, form-finding analyses of a tensegrity
structure and a tensioned membrane structure are illustrated as examples
of minimization problems with constraint conditions.

In section 3, more general cases of static problems of continuum bodies
are taken into account. First, the discrete \textbf{principle of virtual
work}, the \textbf{stationary condition}, and the \textbf{standard
search direction} are formulated as the result of standard procedures.
They can be positioned as the generalizations of those appeared in
section 2 and enables the direct minimization methods feasible on
general cases of static problems of continuum bodies. Finally, some
large deformation analyses of continuum bodies which can be solved
by the \textbf{three-term method} are illustrated.

\section{Two-term method and three-term method}

\subsection{Direct minimization approaches without constraint conditions}

\begin{figure*}[t]
\centering{}\includegraphics{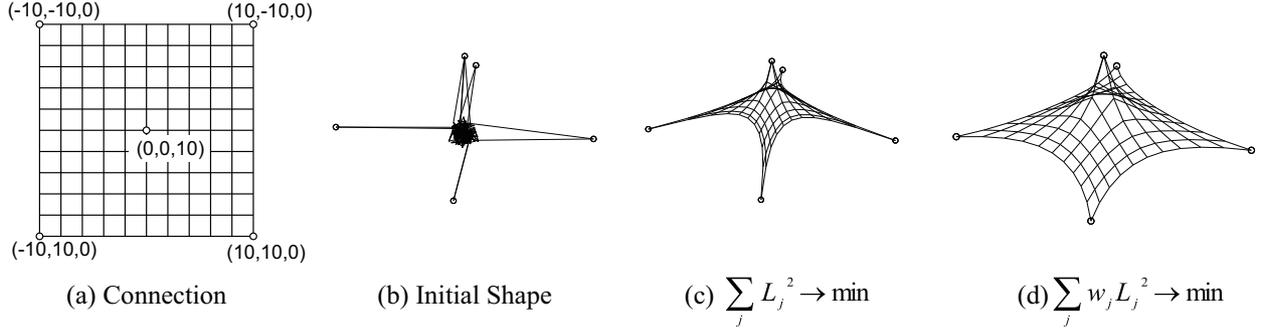}\caption{\label{fig:Form-finding-of-Cable-net}Form-finding of Cable-net Structure}
\end{figure*}

Suppose a form-finding problem of a prestressed cable-net structure
which can be stabilized via introducing prestress (see Fig. \ref{fig:Form-finding-of-Cable-net}).
For example, any solutions of the following stationary problem of
a functional can be used as such a form:

\begin{equation}
\Pi\left(\boldsymbol{x}\right)=\sum_{j}w_{j}L_{j}^{2}\left(\boldsymbol{x}\right)\rightarrow\mathrm{stationary},\label{eq:1}
\end{equation}
where $w_{j},L_{j}$ denote the weight coefficient and the length
of the $j$-th cable respectively. The weight coefficients are the
parameters which are assigned with the aim of varying the form by
varying them, and they are treated as constant in the following formulations.
In addition, $\boldsymbol{x}$ is a column vector which contains the
unknown variables $\left\{ x_{1},\cdots,x_{n}\right\} $.

In this work, $\boldsymbol{x}$ and corresponding gradient vector
are always arranged as 
\begin{equation}
\boldsymbol{x}\equiv\left[\begin{array}{ccc}
x_{1} & \cdots & x_{n}\end{array}\right]^{T},\,\,\mathrm{and}\,\nabla f\equiv\left[\begin{array}{ccc}
\frac{\partial f}{\partial x_{1}} & \cdots & \frac{\partial f}{\partial x_{n}}\end{array}\right].
\end{equation}

By the authors, it has been pointed out \citep{miki2010} that solving
Eq.\eqref{eq:1} by the direct minimization methods can be positioned
as that an equilibrium equation provided by the force density method
\citep{Scheck1974} is represented in a different manner firstly,
and then the equilibrium equation is solved by a direct minimization
method which differ from the method proposed in the original force
density method.

In this work, the unknown variables $\left\{ x_{1},\cdots,x_{n}\right\} $
are always assumed as denoting the Cartesian coordinates of the free
nodes. In addition, remark that those of the fixed nodes are eliminated
beforehand from $\boldsymbol{x}$ and they are directly substituted
into each $L_{j}$.

When Eq.\eqref{eq:1} is solved by the direct minimization methods,
$\Pi\left(\boldsymbol{x}\right)$ is usually called the objective
function. Additionally, the direction of greatest rate of increase
of $\Pi$, namely 
\begin{equation}
\boldsymbol{r}=\frac{\nabla\Pi^{T}}{\left|\nabla\Pi\right|}\label{eq:3}
\end{equation}
is usually adopted as the \textbf{standard search direction}.

The stationary condition of Eq.\eqref{eq:1} is as follows:\textbf{
\begin{equation}
\nabla\Pi=\boldsymbol{0}\Leftrightarrow\sum_{j}2w_{j}L_{j}\nabla L_{j}=\boldsymbol{0}.\label{eq:4}
\end{equation}
}Here, taking the inner product of Eq. \eqref{eq:4} with arbitrary
column vector $\delta\boldsymbol{x}$, namely
\begin{equation}
\delta\boldsymbol{x}=\left[\begin{array}{ccc}
\delta x_{1} & \cdots & \delta x_{n}\end{array}\right]^{T},
\end{equation}
the \textbf{principle of virtual work} can be obtained as: 
\begin{equation}
\delta w=\sum_{j}2w_{j}L_{j}\delta L_{j}=0,\label{eq:6}
\end{equation}
or the \textbf{variational principle} can be found has
\begin{equation}
\delta\Pi=0,\label{eq:7}
\end{equation}
where 
\begin{equation}
\delta f\equiv\nabla f\cdot\delta\boldsymbol{x}
\end{equation}
is the variation of $f$. Due to the arbitrariness of $\delta\boldsymbol{x}$,
Eq. \eqref{eq:6} and Eq. \eqref{eq:7} are always equivalent with
Eq. \textbf{\ref{eq:4}}. It is amazing that the common frameworks
which are provided by the classic mechanics, such as the \textbf{principle
of virtual work} and the \textbf{variational principle}, can be even
found in such a minor force density method.

On the other hand, the \textbf{principle of virtual work} for self-equilibrium
cable-net structures can be expressed as

\begin{equation}
\delta w=\sum_{j}n_{j}\delta L_{j}=0,\label{eq:9}
\end{equation}
where $n_{j}$ denotes the tension of $j$-th cable. By comparing
Eq. \eqref{eq:6} and Eq. \eqref{eq:9}, when $\Pi\left(\boldsymbol{x}\right)$
is stationary, at least one self-equilibrium state can be found as
\begin{equation}
\left[\begin{array}{ccc}
n_{1} & \cdots & n_{m}\end{array}\right]=\left[\begin{array}{ccc}
2w_{1}L_{1} & \cdots & 2w_{m}L_{m}\end{array}\right].
\end{equation}
Therefore, any solution of Eq. \eqref{eq:1} can be used as a form
of cable-net structures that can be prestressed.

When the \textbf{standard search direction} is given by Eq. \eqref{eq:3},
one of the simplest recursive direct minimization methods is given
by
\begin{eqnarray}
\boldsymbol{r}_{\mathrm{Current}} & = & \frac{\nabla\Pi^{T}}{\left|\nabla\Pi\right|}\circ\left(\boldsymbol{x}=\boldsymbol{x}_{\mathrm{Current}}\right),\nonumber \\
\boldsymbol{x}_{\mathrm{Next}} & = & \boldsymbol{x}_{\mathrm{Current}}-\alpha\boldsymbol{r}_{\mathrm{Current}},\label{eq:11}
\end{eqnarray}
which is called the \textbf{two-term method} in this work. Here, ''Current''
and ''Next'' are the current and the next step numbers. As is immediately
noticed, the \textbf{two-term method} is basically identical with
the steepest decent method. The main differences are as follows:
\begin{itemize}
\item The \textbf{standard search direction} is always normalized.
\item Step-size factor $\alpha$ is a parameter which is assigned with the
aims of adjusting the rate of convergence and treated as constant
in the formulations. (In the steepest decent method, step-size is
usually determined by line-search algorithm )
\end{itemize}
The aim of the normalization of the \textbf{standard search direction}
is to prevent the divergence of the computation. Moreover, without
using some computational algorithms to determine $\alpha$ , if $\alpha$
is treated as constant in the formulation and to be adjusted by somebody
via GUI, appropriate $\alpha$ can be found easily. Actually, it was
really easy and intuitive operation to determine $\alpha$ via GUI.

By the way, the rate of global convergence efficiency of the \textbf{two-term
method} is not basically good, as of the steepest decent method usually
is. Because it is supposed that the computation would usually start
from a point which places far from the exact solution, the rate of
global convergence efficiency must be improved. Then, the following
remedy of the \textbf{two-term method} sometimes provides a remarkable
improvement of global convergence efficiency:

\begin{eqnarray}
\boldsymbol{r}_{\mathrm{Current}} & = & \frac{\nabla\Pi^{T}}{\left|\nabla\Pi\right|}\circ\left(\boldsymbol{x}=\boldsymbol{x}_{\mathrm{Current}}\right),\nonumber \\
\boldsymbol{q}_{\mathrm{Next}} & = & 0.98\boldsymbol{q}_{\mathrm{Current}}-\alpha\boldsymbol{r}_{\mathrm{Current}},\nonumber \\
\boldsymbol{x}_{\mathrm{Next}} & = & \boldsymbol{x}_{\mathrm{Current}}+\alpha\boldsymbol{q}_{Next},\label{eq:12}
\end{eqnarray}
which is called the \textbf{three-term method} in this work. When
$\left\{ \boldsymbol{x},\boldsymbol{q},\boldsymbol{r}\right\} $ are
thought as \{position, velocity, acceleration\}, Eq. \eqref{eq:12}
can be positioned as one kind of equation of motion with a damping
term, therefore the basic idea of the \textbf{three-term method} is
almost identical with the dynamic relaxation method\citep{Barnes1999}.
However, as same as in the \textbf{two-term method}, the \textbf{standard
search direction} is also normalized in the \textbf{three-term method}.
Then, it is better to interpret the \textbf{three-term method} as
just one of the recursive direct minimization methods and is not being
based on dynamic mechanics. The factor 0.98 which can be found on
the second line means that 2\% of $\boldsymbol{q}$ is compulsory
cut in each step, which can be interpreted as one kind of damping
factor. This factor is having no basis and being determined by some
experience.

As is mentioned above, because the \textbf{three-term method} is not
based on the dynamics, the following consideration is not precise;
however, the high rate of global convergence efficiency provided by
the \textbf{three-term method} can be understood intuitively when
it is explained with terms of energy conservation law. Namely, due
to the elimination of 2\% of $\boldsymbol{q}$ in each step, the total
energy of the system is compulsory exhausted gradually, then $\Pi$
and $\boldsymbol{q}$ would shortly leach to the minimum value and\textbf{
$\boldsymbol{0}$}.

An numerical model for verification of the \textbf{two-term} and the
\textbf{three-term method} is provided by Fig. \ref{fig:Form-finding-of-Cable-net},
which consists of 5 fixed nodes and 220 tension members. The coordinates
of the fixed nodes are also presented in the figure. As shown by Fig.
\ref{fig:Form-finding-of-Cable-net}(b), initial values of $\left\{ x_{1},\cdots,x_{n}\right\} $
were set by random numbers ranging from -2.5 to 2.5 by the authors,
then it was able to obtain Fig. \ref{fig:Form-finding-of-Cable-net}(c)
and (d)by either \textbf{2-term} or \textbf{3-term method}. Fig. \ref{fig:Form-finding-of-Cable-net}
(c) is the form taking minimum value of the sum of squared length
of all the tension members and the corresponding minimal value was
160.214. Fig. \ref{fig:Form-finding-of-Cable-net} (d) is the form
which was obtained when 4 times greater weight coefficients were assigned
onto the boundary cables and the corresponding minimum value was 188.09.
In both method, 0.2 was used as the step-size factor $\alpha$.

\begin{figure}[!tbh]
\centering{}\includegraphics{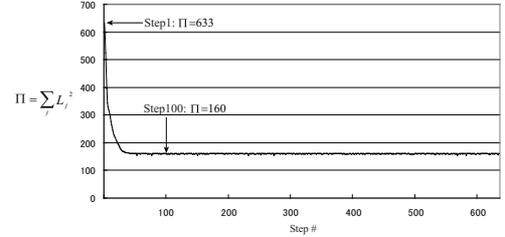}\caption{\label{fig:History-of-()}History of $\Pi$($\alpha=0.2$) (by 3-term
method)}
\end{figure}

\begin{figure}[!tbh]
\centering{}\includegraphics{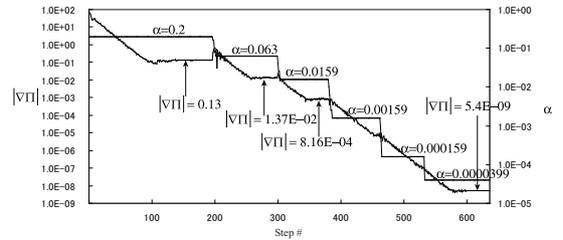}\caption{\label{fig:History-of-}History of $\left|\nabla\Pi\right|$ (by 3-term
method)}
\end{figure}

Fig. \ref{fig:History-of-()} shows the history of the objective function
when Fig. \ref{fig:Form-finding-of-Cable-net} (c) was obtained. As
shown by Fig. \ref{fig:History-of-()}, after a while $\alpha$ was
fixed to 0.2, soon $\Pi$ converged and vibrated around 160. At that
time, the norm of $\left|\nabla\Pi\right|$ was 0.13, which my be
thought as not being sufficiently small. Even such cases, as shown
in Fig. \ref{fig:History-of-}, it is possible to decrease $\left|\nabla\Pi\right|$
gradually by decreasing $\alpha$ gradually. However, this work expects
the \textbf{two-term} and \textbf{three-term method} to be used as
a means of exploring various equilibrium forms by varying the parameters
such as weight coefficients and the coordinates of the fixed nodes
freely, in which $\alpha$ would be kept constant such as 0.2.

\subsection{History of 3-term method}

Here, it must be noted the close relation between Eq. \eqref{eq:12}
and the ''Three-term recursion formulae''. In 1959, ''Three term recursion formulaeh
was firstly presented by M. Engeli, H. Rutishauser et. al \citep{Engeli1959}.
In 1982, M. Papadrakakis stated that the dynamic relaxation method
\citep{Barnes1999} and the conjugate gradient method, can be classified
under the family methods with three-term recursion formulae \citep{Papadrakakis1982}.
Because Eq. \eqref{eq:12} has a common form with the conjugate gradient
method and its basic idea highly resemble the one of the dynamic relaxation
method, it may be possible to position the \textbf{three-term method
}proposed in this work as the simplest method based on the three-term
recursion formulae

\subsection{Direct minimization approaches with constraint conditions}

\begin{figure}[!tbh]
\centering{}\includegraphics{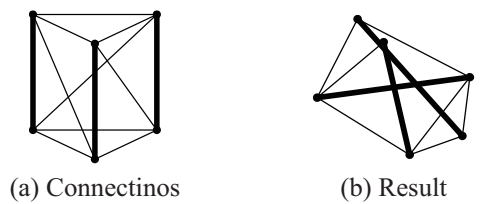}\caption{\label{fig:Form-finding-of-Simplex}Form-finding of \textit{Simplex
Tensegrity}}
\end{figure}

\begin{figure*}[t]
\centering{}%
\begin{tabular}{c>{\centering}p{0.75\textwidth}}
\includegraphics{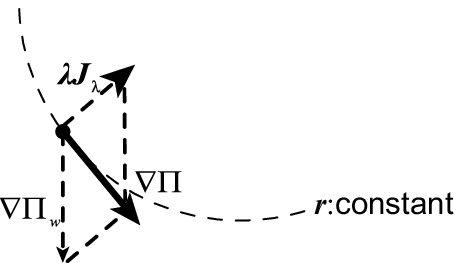} & \includegraphics{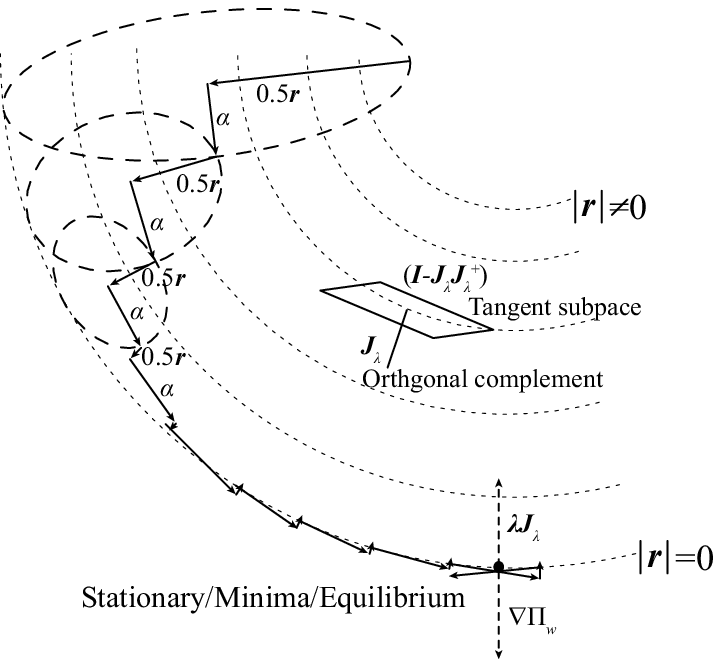}\tabularnewline
(a) Composition of forces & (b) Orthogonal decomposition of search direction\tabularnewline
\end{tabular}\caption{\label{fig:Direct-Minimization-under}Direct Minimization Approaches
under Constraint Conditions}
\end{figure*}
\begin{figure*}[!t]
\noindent \centering{}\includegraphics{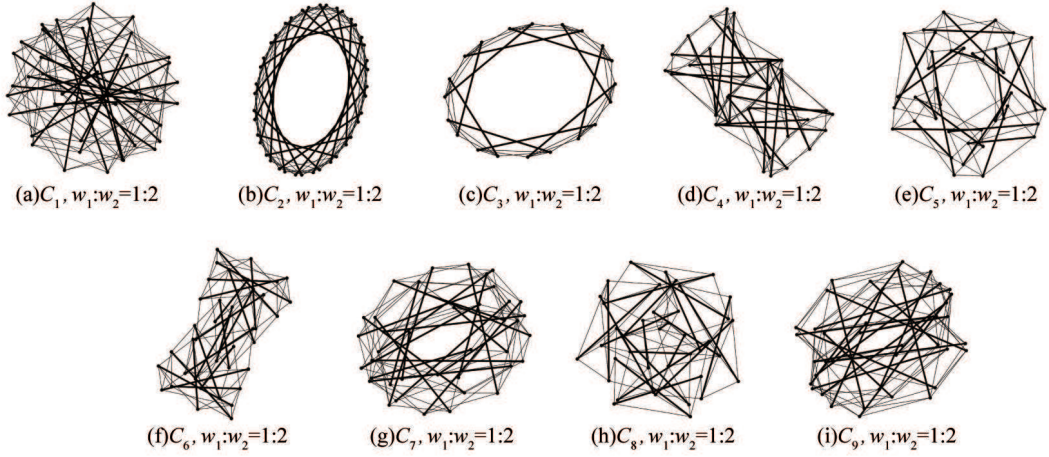}\caption{\label{fig:Form-Finding-of-Tanzbrunnen}Form-Finding of Complicate
Tensegrities}
\end{figure*}
In this section, direct minimization approaches with constraint conditions
are discussed. As an example, let us consider the form-finding problem
of a Simplex Tensegrity structure, which is shown by Fig. \ref{fig:Form-finding-of-Simplex}.
A Simplex Tensegrity is a self-equilibrium structure that consist
of 3 compression members which are shown as thick lines in the figure
and 9 tension members, as thin lines in the figure. In addition, remark
that the members are pin-jointed on only their ends.

In general, it is expected to obtain such a self-equilibrium form
when some objective function with respect to the lengths of the tension
members is minimized by constraining the lengths of compression members.
Then, let us consider the following simple minimization problem with
equally constraint conditions:

\begin{eqnarray}
\Pi_{w}\left(\boldsymbol{x}\right) & = & \sum_{j=1}^{9}w_{j}{L_{j}}^{4}\left(\boldsymbol{x}\right)\rightarrow\min,\label{eq:13}\\
\mathrm{s.\, t.} &  & \begin{cases}
\left(L_{10}-\bar{L}_{10}\right) & =0,\\
 & \vdots\\
\left(L_{12}-\bar{L}_{12}\right) & =0,
\end{cases}\nonumber 
\end{eqnarray}
where $\left\{ L_{1},\cdots,L_{9}\right\} $ denote the lengths of
the tension members and $\left\{ L_{10},\cdots,L_{12}\right\} $ denote
the lengths of the compression members. In addition, $\left\{ w_{1},\cdots,w_{9}\right\} $
are the weight coefficients assigned to every tension members. Moreover,
$\left\{ \bar{L}_{10},\cdots,\bar{L}_{12}\right\} $ are the constraint
values of the lengths of the compression members, which are treated
as constant in the formulations below but are assigned with the aims
of varying the form by varying them. The basis of the power 4 which
is put on $L_{j}$ is not explained in this work, because it has been
already reported by the authors \citep{miki2010}.

Fig. \ref{fig:Form-finding-of-Simplex} (b) shows the form that is
taking the minimum value of $\sum_{j}L_{j}^{4}$ when every lengths
of the compression members are constrained to 10.0. The corresponding
minimum value was 18000 and the methods performed by the authors are
described below.

Applying the \textit{Lagrange} multiplier method, Eq. \eqref{eq:13}
reduces to the following stationary problem of a functional:

\begin{equation}
\Pi(\boldsymbol{x},\boldsymbol{\lambda})=\sum_{j=1}^{m}w_{j}{L_{j}}^{4}\left(\boldsymbol{x}\right)+\sum_{k=1}^{r}\lambda_{k}\left(L_{m+k}\left(\boldsymbol{x}\right)-\bar{L}_{m+k}\right)\rightarrow\mathrm{stationary},\label{eq:stationary_prob.}
\end{equation}
where the first sum is taken for all the tension members and the second
sum is taken for all the compression members. In addition, $\boldsymbol{\lambda}$
is a row vector containing the multipliers $\left\{ \lambda_{1},\cdots,\lambda_{r}\right\} $.
From now on, let $\boldsymbol{x},\boldsymbol{\lambda}$ and the corresponding
gradient vectors be arranged as {\footnotesize 
\begin{eqnarray}
\boldsymbol{x}\equiv\left[\begin{array}{ccc}
x_{1} & \cdots & x_{n}\end{array}\right]^{T} & , & \,\frac{\partial f}{\partial\boldsymbol{x}}\equiv\left[\begin{array}{ccc}
\frac{\partial f}{\partial x_{1}} & \cdots & \frac{\partial f}{\partial x_{n}}\end{array}\right],\\
\mathbf{\boldsymbol{\lambda}}\equiv\left[\begin{array}{ccc}
\lambda_{1} & \cdots & \lambda_{r}\end{array}\right] & , & \,\frac{\partial f}{\partial\boldsymbol{\lambda}}\equiv\left[\begin{array}{ccc}
\frac{\partial f}{\partial\lambda_{1}} & \cdots & \frac{\partial f}{\partial\lambda_{r}}\end{array}\right]^{T}.
\end{eqnarray}
}Then, let the gradient operator $\nabla$ be defined by $\nabla f\equiv\frac{\partial f}{\partial\boldsymbol{x}}$.

The stationary condition of Eq. \eqref{eq:stationary_prob.} can be
expressed as a set of two conditions: 
\begin{equation}
\frac{\partial\Pi}{\partial\boldsymbol{x}}=\boldsymbol{0}\,\,\&\,\,\frac{\partial\Pi}{\partial\boldsymbol{\lambda}}=\boldsymbol{0}.\label{eq:17}
\end{equation}
Thus, in this work, it is important that\emph{\noun{ }}unknown variables
$\boldsymbol{x}$ and multipliers $\boldsymbol{\lambda}$ are explicitly
distinguished and the corresponding stationary conditions are discussed
separately.

First, let us discuss the first stationary condition, namely
\begin{equation}
\nabla\Pi=\boldsymbol{0}\Leftrightarrow\sum_{j}4w_{j}{L_{j}}^{3}\nabla L_{j}+\sum_{k}\lambda_{k}\nabla L_{k}=\boldsymbol{0},\label{eq:stat.cond1}
\end{equation}
and its general form is expressed as 
\begin{equation}
\nabla\Pi=\nabla\Pi_{w}+\boldsymbol{\lambda}\cdot\boldsymbol{J}_{\lambda}=\boldsymbol{0},\label{eq:stat_cond2}
\end{equation}
where $\nabla\Pi_{w}$ is the gradient of the objective function and
$\boldsymbol{J}_{\lambda}$ is an Jacobian matrix given by
\begin{equation}
\boldsymbol{J}_{\lambda}\equiv\left[\begin{array}{c}
\nabla L_{m+1}\\
\vdots\\
\nabla L_{m+r}
\end{array}\right],\label{eq:aaa}
\end{equation}
which must be refreshed in each step.

In this work, the number of the constraint conditions are assumed
as being smaller than the number of the unknown variables. In addition,
bad conditioned problems, such that satisfaction of the constraint
conditions is almost impossible, are not discussed. The simple interpretation
of above statements in terms of mathematics is that $\boldsymbol{J}_{\lambda}$
is always supposed as full-rank and the number of the columns is supposed
as being greater than the one of the rows.

Taking into account the making use of the direct minimization methods
on this problem, the indeterminacy of $\nabla\Pi$ must be solved,
i.e. due to the unknown multipliers $\left\{ \lambda_{1},\cdots,\lambda_{r}\right\} $
, $\nabla\Pi$ can not be determined uniquely, and this makes the
\textbf{two-term }and the \textbf{three-term method} infeasible. In
contrast, if an additional rule is supplemented with the aim of determining
$\left\{ \lambda_{1},\cdots,\lambda_{r}\right\} $ uniquely, $\nabla\Pi$
is also determined uniquely and both the \textbf{two-term }and the
\textbf{three-term method} turn to feasible. One of the simplest ideas
is making use of the Moore-Penrose type pseudo inverse matrix $\boldsymbol{J}_{\lambda}^{+}$.

First, Eq. \eqref{eq:stat_cond2} is transformed into 
\begin{equation}
\boldsymbol{\lambda}\cdot\boldsymbol{J}_{\lambda}=-\nabla\Pi_{w},\label{eq:a-1}
\end{equation}
and then, $\boldsymbol{\lambda}$ can be determined by 
\begin{equation}
\boldsymbol{\lambda}=-\nabla\Pi_{w}\cdot\boldsymbol{J}_{\lambda}^{+},\label{eq:lambda}
\end{equation}
which provides basically a least norm solution. When $\boldsymbol{J}_{\lambda}$
is supposed as fullrank, it is simply given by $\boldsymbol{J}_{\lambda}^{+}=\boldsymbol{J}_{\lambda}^{T}\cdot\left(\boldsymbol{J}_{\lambda}\cdot\boldsymbol{J}_{\lambda}^{T}\right)^{-1}$.
One may feel it is very hard to adopt such a least squared solution
because it is not an exact solution; however, when $\boldsymbol{x}$
turns to a solution, $\boldsymbol{\lambda}$ given by Eq. \eqref{eq:lambda}
turns to a least norm solution and when Eq. \eqref{eq:lambda} gives
a least norm solution, it implies that a stationary point has been
obtained. Otherwise, when Eq. \eqref{eq:lambda} gives a least squared
solution, it implies that $\boldsymbol{x}$ still has not leached
to a stationary point\noun{,} therefore, a supplement of an additional
rule to determine $\boldsymbol{\lambda}$ uniquely must not be interfered
by any reason. 

As the result of above discussion, a unique mapping from $\boldsymbol{x}$
to $\nabla\Pi$ can be defined by
\begin{equation}
\nabla\Pi\equiv\nabla\Pi_{w}+\boldsymbol{\lambda}\cdot\boldsymbol{J}_{\lambda}=\boldsymbol{0}\circ\left(\boldsymbol{\lambda}=-\nabla\Pi_{w}\cdot\boldsymbol{J}_{\lambda}^{+}\right),\label{eq:grad}
\end{equation}
which determines a gradient vector filed and thus both the \textbf{two-term}
and the \textbf{three-term method} turn to feasible. The determination
of $\left\{ \nabla\Pi,\boldsymbol{\lambda}\right\} $ by using Eq.
\eqref{eq:grad} is essentially identical with the \textbf{dual estimate},
which is defined in linear programming theory, particularly in the
context of the primal affine scaling method \citep{Dikin1967}.

By the way, the substitution appeared in Eq. \eqref{eq:grad} can
be performed immediately, and then Eq. \eqref{eq:grad} reduces to
\begin{equation}
\nabla\Pi=\nabla\Pi_{w}\cdot(\boldsymbol{I}-\boldsymbol{J}_{\lambda}\cdot{\boldsymbol{J}_{\lambda}}^{+}),\label{eq:projected_gradient}
\end{equation}
which is widely known as the projected gradient in terms of the projected
gradient method and in which $\boldsymbol{\lambda}$ is eliminated.
However, the multipliers are always calculated explicitly in this
work because the \textbf{dual estimate} can be interpreted to the
\textbf{composition of forces} when $\nabla\Pi_{w}$ is considered
as a force, $\boldsymbol{\lambda}$ as a reaction force and $\nabla\Pi$
as a resultant force as shown in Fig. \ref{fig:Direct-Minimization-under}
(a).

Let us recall and discuss the second stationary condition, namely
\begin{equation}
\frac{\partial\Pi}{\partial\boldsymbol{\lambda}}=\boldsymbol{0}\Leftrightarrow\begin{cases}
\left(L_{m+1}-\bar{L}_{m+1}\right) & =0\\
 & \vdots\\
\left(L_{m+r}-\bar{L}_{m+r}\right) & =0
\end{cases},\label{eq:conditions}
\end{equation}
which is apparently the prescribed equally constraint conditions themselves.
One of the simplest ideas to satisfy Eq. \eqref{eq:conditions} is
to solve simultaneous linear equations such as

\begin{equation}
\boldsymbol{J}_{\lambda}\cdot\Delta\boldsymbol{x}=-\boldsymbol{r},
\end{equation}
where $\Delta\boldsymbol{x}$ is a correction vector of $\boldsymbol{x}$
and $\boldsymbol{r}$ is a residual vector given by
\begin{equation}
\boldsymbol{r}=\left[\begin{array}{c}
L_{m+1}\left(\boldsymbol{x}\right)-\bar{L}_{m+1}\\
\vdots\\
L_{m+r}\left(\boldsymbol{x}\right)-\bar{L}_{m+r}
\end{array}\right].
\end{equation}
The definition of $\boldsymbol{J}_{\lambda}$ is apparently identical
with Eq. \eqref{eq:aaa}, but should be refreshed again. Here, the
Moore-Penrose type pseudo inverse $\boldsymbol{J}_{\lambda}^{+}$
plays an important role again to determine $\Delta\boldsymbol{x}$
as

\begin{equation}
\Delta\boldsymbol{x}=-{\boldsymbol{J}_{\lambda}}^{+}\cdot\boldsymbol{r},\label{eq:28}
\end{equation}
which basically gives a least norm solution. In addition, because
$\boldsymbol{x}$ can places far from the hyper-surface on which the
constraint conditions are satisfied, it is highly recommended to rescale
$\Delta\boldsymbol{x}$ to prevent the computation being unstable,
such as

\begin{equation}
\boldsymbol{x}_{\mathrm{Current}}:=\boldsymbol{x}_{\mathrm{Current}}+0.5\Delta\boldsymbol{x},\label{eq:Residual}
\end{equation}
where : symbol represents a substitution of the right hand side into
left hand side. If Eq. \eqref{eq:Residual} is always performed once
after the execution of Eq. \eqref{eq:11} or Eq. \eqref{eq:12} in
each step, $\boldsymbol{x}$ would gradually approaches to the hyper-surface
on which the prescribed constraint conditions are satisfied and soon,
the motion of $\boldsymbol{x}$ generated by the \textbf{two-term}
or the \textbf{three-term method} will be constrained onto such a
hyper-surface, as shown by Fig. \ref{fig:Direct-Minimization-under}(b).
By using either \textbf{two-term} or \textbf{three-term method}, Fig.
\ref{fig:Form-finding-of-Simplex} (b) was obtained. By introducing
Eq. \eqref{eq:Residual}, it is also enabled starting the computation
from random numbers. As same as in the previous section, to obtain
Fig. \ref{fig:Form-finding-of-Simplex} (b), the authors gave random
numbers ranging from -2.5 to 2.5 to the initial values of $\left\{ x_{1},\cdots,x_{n}\right\} $
and set the step-size factor $\alpha$ as 0.2.

By comparing Eq. \eqref{eq:projected_gradient} and Eq. \eqref{eq:28},
one may notice that $\nabla\Pi$ and $\Delta\boldsymbol{x}$ are row
and column vectors which are selected from completely decomposed two
spaces that are orthogonal to each other, because $\left(\boldsymbol{I}-\boldsymbol{J}_{\lambda}^{+}\cdot\boldsymbol{J}_{\lambda}\right)$
represents the kernel of $\boldsymbol{J}_{\lambda}^{+}$ and vice
versa. As is depicted in Fig. \ref{fig:Direct-Minimization-under}
(b), the $n$-dimensional search space (usually assumed as an Euclidean
space) that $\boldsymbol{x}$ belongs to is firstly decomposed into
a group of hyper-surfaces on which residual vector $\boldsymbol{r}$
taking the same value. Then, on each point of each hyper-surface,
the vector space attached to each point is completely decomposed into
the tangent subspace and the orthogonal complement. Finally, each
of $\nabla\Pi$ and $\Delta\boldsymbol{x}$ is selected from each
of the tangent subspace and the orthogonal complement respectively.
Thus, the feature of the method proposed in this work which should
be emphasized is that two subspaces that are orthogonal to each other
correspond to two separated stationary conditions, and, two different
strategies are performed independently on each subspace.

By the way, when minimization problems with constraint conditions
are solved by the method proposed above, the \textbf{two-term method}
become sometimes useless, particularly on complicate problems. In
contrast, by using the \textbf{three-term method}, it was still possible
to find the forms of complicate structures such as the tensegrities
shown by Fig. \ref{fig:Form-Finding-of-Tanzbrunnen} (see \citep{miki2010}).

\begin{figure*}[!tb]
\noindent \centering{}\includegraphics{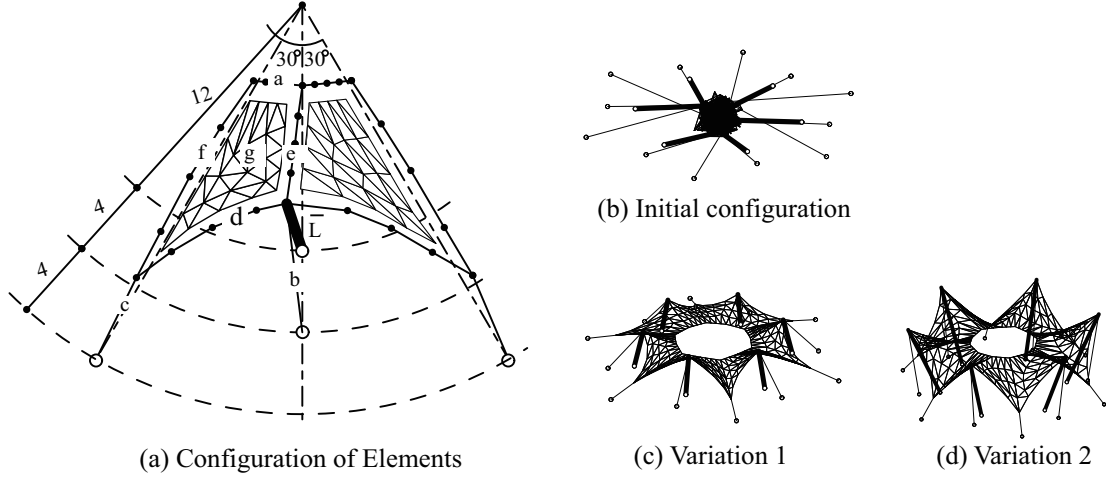}\caption{\label{fig:Form-Finding-of-Tanzbrunnen-1}Form-Finding of Tanzbrunnen
in Cologne (F. Otto, 1959)}
\end{figure*}

Such complicate problem on which the \textbf{three-term method} works
better than the \textbf{two-term method} can be easily found widely.
Fig. \ref{fig:Form-Finding-of-Tanzbrunnen-1} shows such another form-finding
analysis, in which, the analytical model consists of cables, membranes,
compression members and fixed points, and based on the famous Tanzbrunnen
in Cologne iFrei Otto , 1959j. The selected stationary problem that
was solved is as follows:
\begin{align}
\Pi\left(\boldsymbol{x},\boldsymbol{\lambda}\right) & =\sum_{j=1}^{m}w_{j}L_{j}^{4}\left(\boldsymbol{x}\right)+\sum_{k}w_{k}S_{k}^{2}\left(\boldsymbol{x}\right)\nonumber \\
 & +\sum_{l=1}^{r}\lambda_{l}\left(L_{m+l}\left(\boldsymbol{x}\right)-\bar{L}_{m+l}\right)\rightarrow\mathrm{stationary},\label{eq:30}
\end{align}
where the cables, the membranes are subdivided into line elements
and triangle elements, and the first sum is taken for all the line
elements and the second sum is taken for all the triangle elements.
In addition, the third sum is taken for all the compression members,
which can be composed into a stationary problem by applying the \emph{Lagrange}
multiplier method to the constraint conditions. Moreover, $L_{j}$,
$S_{k}$, and $L_{m+l}$ are the functions that respectively represent
the length of a line element, the area of a triangle element, and
the length of a compression member.

The statinoary condition with respect to $\boldsymbol{x}$ can be
expressed as :

\begin{equation}
\nabla\Pi=\sum_{j}^{m}4w_{j}L_{j}^{3}\nabla L_{j}+\sum_{k}2w_{k}S_{k}\nabla S_{k}+\sum_{l}^{r}\lambda_{l}\nabla L_{m+l}=\boldsymbol{0},\label{eq:31}
\end{equation}
and then, taking the inner-product between $\delta\boldsymbol{x}$
and Eq. \eqref{eq:31}, the \textbf{principle of virtual work} for
this problem can be expressed as: 
\begin{equation}
\delta w=\sum_{j}^{m}4w_{j}L_{j}^{3}\delta L_{j}+\sum_{k}2w_{k}S_{k}\delta S_{k}+\sum_{l}^{r}\lambda_{l}\delta L_{m+l}=0.
\end{equation}
As means of solving Eq. \eqref{eq:30}, while \textbf{two-term method}
was completely useless, the \textbf{three-term method} worked really
fine. What is more important is that it was also possible to vary
the form by varying the weight coefficients or the lengths of the
compression members and to explore the possible self-equilibrium forms.

On the basis of above considerations, this research is strongly focused
on the \textbf{three-term method,} even though the \textbf{two-term
method}, or the steepest decent method, is sometimes described as
one of the most standard direct minimization methods. In the next
section, the formulations for solving various types of static problems
of continuum bodies by the \textbf{three-term method} are presented.

\noindent 

\section{Continuum mechanics}

When the gradient of the volume of a tetrahedron element, $\nabla V_{j}$,
is added to a set of $\nabla L_{j}$ and $\nabla S_{j}$, a compact
framework in which a set of $\left\{ \nabla L_{j},\nabla S_{j},\nabla V_{j}\right\} $
is adopted as basic gradient vectors can be formed; however, this
framework is almost useless, because, while the stress tensor defined
on a 3-dimensional body usually has 6 degree of freedom, the degree
of freedom that $\nabla V_{j}$ has is at most only the scalar multiplication.
Hence, a further consideration on the continuum mechanics must be
needed, if taking into account the \textbf{three-term method} on solving
various types of static problems of continuum mechanics. In this section,
the discrete \textbf{principle of virtual work}, the \textbf{stationary
condition}, the \textbf{standard search direction} are derived from
the \textbf{principle of virtual work} that is a governing field equation.
Additionally, a general form of $\left\{ \nabla L_{j},\nabla S_{j},\nabla V_{j}\right\} $
is presented as $\boldsymbol{\omega}_{j}^{N}\left(T_{\cdot k}^{i}\right)$.

\subsection{Minimal surfaces and uniform stress surfaces}

From now on, \emph{Einstein} summation convention is used. In this
subsection, the relation between minimal surfaces and the uniform
stress surfaces are discussed. A minimal surface is a surface such
that when its form is varied arbitrarily by fixing its boundary, its
surface area does not change.

In general, the surface area of a surface having a fixed boundary
can be expressed as:

\begin{equation}
a\equiv\int_{a}\mathrm{da},\,\,\mathrm{da}\equiv\sqrt{\det g_{ij}}d\theta^{1}d\theta^{2}\,\,\left(1\le i,j\le2\right),
\end{equation}
where $g_{ij},\,\theta^{1},\,\theta^{2}$ represents the \emph{Riemannian}
metric and the local coordinate parameters which are defined on each
point of the surface. Then, the variation of surface area can be expressed
as:

\begin{equation}
\delta a=\int_{a}\delta\sqrt{\det g_{ij}}d\theta^{1}d\theta^{2}.\label{eq:40}
\end{equation}
Here, it is widely known that $\delta\sqrt{\det g_{ij}}$ can be expanded
as
\begin{equation}
\delta\sqrt{\det g_{ij}}=\frac{1}{2}g^{ij}\delta g_{ij},\label{eq:41}
\end{equation}
where $g^{ij}$ is defined as the inverse of $g_{ij}$, namely $g^{ij}=\left(g_{ij}\right)^{-1}$.
Additionally, $\delta g_{ij}$ is not completely arbitrary but must
be geometrically admissible, namely $\delta g_{ij}$ must be expressed
as

\begin{equation}
\delta g_{ij}=\left(\nabla_{i}\delta u^{k}\right)g_{kj}+\left(\nabla_{j}\delta u^{k}\right)g_{ki}-2h_{ij}\delta u^{3},\label{eq:33}
\end{equation}
 where $\left\{ \delta u^{1},\delta u^{2},\delta u^{3}\right\} $
are arbitrary scalar fields such that each $\delta u^{k}$ satisfies
$\delta u^{k}=0$ on the boundary; however the detail of Eq. \eqref{eq:33}
is not discussed in this work because what is only needed for the
\textbf{three-term method} is just an approximation of $\delta g_{ij}$
and not $\delta g_{ij}$ itself. While Eq. \eqref{eq:33} is a field
equation, the aim of this work is to avoid such complicate and difficult
field equations and to obtain an approximated solution easily by the
direct minimization methods.

Substituting Eq. \eqref{eq:41} into Eq. \eqref{eq:40}, the \textbf{minimal
surface problem} can be expressed as:
\begin{equation}
\delta a=0\Leftrightarrow\frac{1}{2}\int_{a}g^{ij}\delta g_{ij}\mathrm{da}=0.\label{eq:43}
\end{equation}
Eq. \eqref{eq:43} has a close relation with the \textbf{principle
of virtual work} of self-equilibrium membranes whose boundary is fixed:
\begin{equation}
\delta w=\frac{1}{2}\int_{a}t\sigma^{ij}\delta g_{ij}\mathrm{da}=0\,\,\left(1\le i,j\le2\right),
\end{equation}
where $t\,\mathrm{and}\,\sigma^{ij}$ respectively denote the thickness
and the Cauchy stress tensor defined on each point of the surface.
Additionally, $\frac{1}{2}\delta g_{ij}$ is used instead of the variation
of strain due to the essential identity between them.

Using raising and lowering indices law of tensors, i.e. $X^{ij}=X_{\cdot k}^{i}g^{kj}$,
the \textbf{principle of virtual work} is transformed into: 
\begin{equation}
\delta w=\frac{1}{2}\int_{a}t\sigma_{\cdot k}^{i}g^{kj}\delta g_{ij}\mathrm{da}=0\,\,\left(1\le i,j,k\le2\right).\label{eq:45}
\end{equation}
Moreover, when a new stress tensor $T_{\cdot k}^{i}$ is defined by
\begin{equation}
T_{\cdot k}^{i}\equiv t\sigma_{\cdot k}^{i},
\end{equation}
Eq. \eqref{eq:45} is transformed again into:
\begin{equation}
\delta w=\frac{1}{2}\int_{a}T_{\cdot k}^{i}g^{kj}\delta g_{ij}\mathrm{da}=0\,\,\left(1\le i,j,k\le2\right).
\end{equation}
On the other hand, Eq. \eqref{eq:43}, the \textbf{minimal surface
problem} is also transformed into: 
\begin{equation}
\delta a=\frac{1}{2}\int_{a}\delta_{\cdot j}^{i}g^{kj}\delta g_{ij}\mathrm{da}=0\,\,\left(1\le i,j,k\le2\right),
\end{equation}
which can be a simple demonstration of the essential identity between
minimal surfaces and uniform stress surfaces.

\subsection{Principle of virtual work for \textit{N}-Dimensional \textit{Riemannian}
manifolds}

In this subsection, the formulations appeared in the previous subsection
are generalized into any dimensional spaces from 2-dimensional spaces
(surfaces). The length, the area, and the volume of a curve, a surface,
and a body which have a boundary are expressed as:

\begin{equation}
l\equiv\int_{l}\mathrm{dl},\,\, a\equiv\int_{a}\mathrm{da},\,\, v\equiv\int_{v}\mathrm{dv},\label{eq:49}
\end{equation}
where $\mathrm{dl},\mathrm{da},\mathrm{dv}$ are respectively called
the line element, the surface element, and the volume element which
are defined by
\begin{eqnarray}
\mathrm{dl} & \equiv & \sqrt{g_{11}}d\theta^{1},\label{eq:50}\\
\mathrm{da} & \equiv & \sqrt{\det g_{ij}}d\theta^{1}d\theta^{2}\left(1\le i,j\le2\right),\label{eq:51}\\
\mathrm{dv} & \equiv & \sqrt{\det g_{ij}}d\theta^{1}d\theta^{2}d\theta^{3}\left(1\le i,j\le3\right),\label{eq:52}
\end{eqnarray}
where $g_{ij}$ represent the \emph{Riemannian} metrices defined on
each point of each geometry. Such geometries on which \emph{Riemannian}
metric is defined on each point, can be classified as 1,2,3-dimensional
\emph{Riemannian} manifold.

Noticing the common forms appeared in Eq. \eqref{eq:49}, \eqref{eq:50},
\eqref{eq:51}, \eqref{eq:52}\emph{,} it is very natural to define
the volume element and the volumen of a \emph{N-}dimensional \emph{Riemannian}
manifold \emph{M} by
\begin{equation}
\mathrm{dv}^{N}\equiv\sqrt{\det g_{ij}}d\theta^{1}\cdots d\theta^{N},\,\, v^{N}\equiv\int_{M}\mathrm{dv}^{N},
\end{equation}
then the variation of the volume of \emph{M} can be expressed by 
\begin{equation}
\delta v^{N}=\frac{1}{2}\int_{M}g^{ij}\delta g_{ij}\mathrm{dv}{}^{N}.
\end{equation}
Hence, the\textbf{ minimal volume problem} of \emph{M} can be defined
by:
\begin{equation}
\delta v^{N}=\frac{1}{2}\int_{M}g^{ij}\delta g_{ij}\mathrm{dv}{}^{N}=0\,\,\left(1\le i,j\le N\right).\label{eq:57-1}
\end{equation}
By the way, the self-equilibrium equations of cables, membranes, and
3-dimensional bodies whose boundary is fixed can be expressed in the
form of the \textbf{principle of virtual work} as follows: 
\begin{equation}
\delta w^{1}=\frac{1}{2}\int_{l}A\sigma_{\cdot k}^{i}g^{kj}\delta g_{ij}\mathrm{dl}=0\,\,\left(i,j,k=1\right),\label{eq:56}
\end{equation}
\begin{equation}
\delta w^{2}=\frac{1}{2}\int_{a}t\sigma_{\cdot k}^{i}g^{kj}\delta g_{ij}\mathrm{da}=0\,\,\left(1\le i,j,k\le2\right),\label{eq:57}
\end{equation}
\begin{equation}
\delta w^{3}=\frac{1}{2}\int_{v}\sigma_{\cdot k}^{i}g^{kj}\delta g_{ij}\mathrm{dv}=0\,\,\left(1\le i,j,k\le3\right),\label{eq:58}
\end{equation}
where $t$ and $A$ respectively denote the sectional area of a cable
and the thickness of a membrane.

Here, when new stress tensor $T_{\cdot k}^{i}$ is defined for each
dimension individually as: \textbf{
\begin{equation}
T_{\cdot k}^{i}\equiv A\sigma_{\cdot k}^{i}\left(N=1\right),\, T_{\cdot k}^{i}\equiv t\sigma_{\cdot k}^{i}\left(N=2\right),\,\mathrm{and}\, T_{\cdot k}^{i}=\sigma_{\cdot k}^{i}\left(N=3\right),
\end{equation}
}Eq. \eqref{eq:56}, Eq. \eqref{eq:57-1} and Eq. \eqref{eq:58} are
unified into:
\begin{equation}
\delta w^{N}=\frac{1}{2}\int_{M}T_{\cdot k}^{i}g^{kj}\delta g_{ij}\mathrm{dv}^{N}=0\,\,\left(1\le i,j,k\le N\right),\label{eq:60}
\end{equation}
which is the \textbf{principle of virtual work} for self-equilibrium
\emph{N}-dimensional \emph{Riemannian} manifold \emph{M}.

Here, Eq. \eqref{eq:57-1}, the \textbf{minimal volume problem} of
\emph{M}, can be transformed into: 
\begin{equation}
\delta v^{N}=\frac{1}{2}\int_{M}\delta_{\cdot k}^{i}g^{ij}\delta g_{ij}\mathrm{dv}^{N}=0\,\,\left(1\le i,j,k\le N\right).\label{eq:61}
\end{equation}
By comparing Eq. \eqref{eq:60} and Eq. \eqref{eq:61}, it can be
noticed that the\textbf{ minimal volume problem} is a special cases
of the \textbf{principle of virtual work} such that $T_{\cdot k}^{i}=\delta_{\cdot k}^{i}$,
and the \textbf{principle of virtual work} is one of the natural generalizations
of the \textbf{minimal volume problem}. In general, $\delta_{\cdot k}^{i}$
can be classed with the unit matrix.

\subsection{Galerkin method}

The \textbf{principle of virtual work} which is defined in the previous
sub section, i.e.
\begin{equation}
\delta w=\frac{1}{2}\int_{M}T_{\cdot k}^{i}g^{kj}\delta g_{ij}\mathrm{dv}^{N}=0\label{eq:pvw-1}
\end{equation}
is basically a field equation; namely the degree of freedom of $\delta g_{ij}$
is infinite. Then, with the aim of solving the \textbf{principle of
virtual work} by the direct minimization methods, in this subsection,
discrete \textbf{principle of virtual work} is deduced.

First, when the form is explicitly represented by $n$ independent
parameters such as $\left\{ x_{1},\cdots,x_{n}\right\} $, then $\boldsymbol{x},\delta\boldsymbol{x},\,\mathrm{and}\,\nabla f=\frac{\partial f}{\partial\boldsymbol{x}}$
can be defined with the same manner of section 2. When, the degree
of freedom of the form is $n$, then at most $n$ independent $\delta g_{ij}$
can satisfy Eq. \eqref{eq:pvw-1}. Thus, in general, any form on which
$n$ independent $\delta g_{ij}$ can satisfy Eq. \eqref{eq:pvw-1}
is usually adopted as an approximated solution. One of such natural
ways of giving $\delta g_{ij}$ is altering $\delta g_{ij}$ into
\begin{equation}
\delta\tilde{g}_{ij}=\nabla g_{ij}\cdot\delta\boldsymbol{x},\label{eq:64}
\end{equation}
which is essentially the \emph{Galerkin} method. If $\delta g_{ij}$
is altered into Eq. \eqref{eq:64}, the discrete \textbf{principle
of virtual work} (weak form) is obtained as:

\begin{equation}
\delta w^{N}=\frac{1}{2}\int_{M}T_{\cdot k}^{i}g^{kj}\left(\nabla g_{ij}\cdot\delta\boldsymbol{x}\right)\mathrm{dv}^{N}=0,
\end{equation}
then, letting $\delta\boldsymbol{x}$ out of the integral operator,
discrete \textbf{principle of virtual work} (strong form) is obtained
as: 
\begin{equation}
\Leftrightarrow\left(\frac{1}{2}\int_{M}T_{\cdot k}^{i}g^{kj}\nabla g_{ij}\mathrm{dv}^{N}\right)\cdot\delta\boldsymbol{x}=0,
\end{equation}
finally, due to the arbitrariness of $\delta\boldsymbol{x}$,
\begin{equation}
\Leftrightarrow\boldsymbol{\omega}=\frac{1}{2}\int_{M}T_{\cdot k}^{i}g^{kj}\nabla g_{ij}\mathrm{dv}^{N}=\boldsymbol{0},
\end{equation}
which is the discrete \textbf{stationary condition} and can be also
positioned as a discrete form of a self-equilibrium equation.

When external forces are acting on the manifold $M$, the discrete
\textbf{principle of virtual work} (strong form) is firstly expressed
as

\textbf{
\begin{equation}
\left(\frac{1}{2}\int_{M}T_{\cdot k}^{i}g^{kj}\nabla g_{ij}\mathrm{dv}^{N}\right)\cdot\delta\boldsymbol{x}=\boldsymbol{p}\cdot\delta\boldsymbol{x},
\end{equation}
}and it follows
\begin{equation}
\Leftrightarrow\boldsymbol{\omega}=\frac{1}{2}\int_{M}T_{\cdot k}^{i}g^{kj}\nabla g_{ij}\mathrm{dv}^{N}-\boldsymbol{p}=\boldsymbol{0},
\end{equation}
where $\boldsymbol{p}$ is a row vector containing the components
of the nodal loads, which should be basically derived via some discretization
process of continuum load but further detail is not discussed in this
work because it has been already discussed in the usual finite element
formulations.

Then, since the discrete s\textbf{tationary condition} is an $n$-order
simultaneous non-linear equations and the number of the unknown variables
is $n$ so that basically it can be solved. In addition, when the
discrete \textbf{stationary condition} is solved by the direct minimization
methods, 

\begin{equation}
\boldsymbol{r}=\frac{\boldsymbol{\omega}^{T}}{\left|\boldsymbol{\omega}\right|},
\end{equation}
is adopted as the \textbf{standard search direction}.

\subsection{$N$-dimensional Simplex elements}

\begin{figure*}[!tbh]
\centering{}\includegraphics{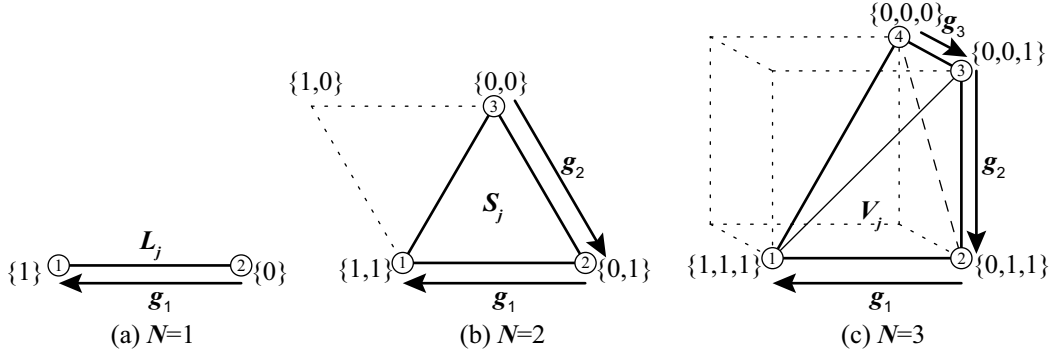}\caption{\label{fig:Simplex-Elements}Simplex Elements}
\end{figure*}

The discrete stationary condition $\boldsymbol{\omega}=\boldsymbol{0}$
which is derived in the previous subsection still contains integral
operator, which is the last obstacle to be overcome. In this subsection,
as a powerful means of calculating $\boldsymbol{\omega}$ on general
numerical environment, \emph{N}-dimensional Simplex element is presented.

When the integral domain is subdivided into $m$ elements, if element
integral is defined by
\begin{equation}
\boldsymbol{\omega}_{j}^{N}\left(T_{\cdot\gamma}^{\alpha}\right)\equiv\frac{1}{2}\int_{j}T_{\cdot\gamma}^{\alpha}g^{\gamma\beta}\nabla g_{\alpha\beta}\mathrm{dv}^{N},\label{eq:71}
\end{equation}
where the integral operation is calculated separately within each
element, then, $\boldsymbol{\omega}$ can be simply expressed as\textbf{
\begin{equation}
\boldsymbol{\omega}=\sum_{j}\boldsymbol{\omega}_{j}^{N}\,\,\mathrm{or}\,\,\boldsymbol{\omega}=\sum_{j}\boldsymbol{\omega}_{j}^{N}-\boldsymbol{p}.
\end{equation}
}The most simplest idea to calculate Eq. \eqref{eq:71} is to let
the integrated function constant within each element. Fig. \eqref{fig:Simplex-Elements}
shows 1,2,3-dimensional Simplex elements, which are apparently just
the line, triangle, and tetrahedron element having $N+1$ nodes.

From now on, first, let $\left\{ \boldsymbol{p}_{1},\cdots,\boldsymbol{p}_{N+1}\right\} $
be a set of the Cartesian coordinates of the nodes of an element,
and then second, let $\left\{ \theta^{1},\cdots,\theta^{N}\right\} $
be a simple local coordinate defined on the element. Third, let each
coordinate parameter be taking the value from 0 to 1. Then, finally,
the global coordinate (assumed as the Cartesian coordinate) of each
point within the element can be given by an interpolation function
defined by

{\footnotesize 
\begin{equation}
\boldsymbol{r}\left(\theta^{1},\cdots,\theta^{N}\right)=\theta^{1}\left(\boldsymbol{p}_{1}-\boldsymbol{p}_{2}\right)+\cdots+\theta^{N}\left(\boldsymbol{p}_{N}-\boldsymbol{p}_{N+1}\right)+\boldsymbol{p}_{N+1}.
\end{equation}
}Then, referring to the definition of the base vectors, namely 
\begin{equation}
\boldsymbol{g}_{i}\equiv\frac{\partial\boldsymbol{r}}{\partial\theta^{i}},\label{eq:71-1}
\end{equation}
$\boldsymbol{g}_{1}\cdots\boldsymbol{g}_{N}$ can be calculated by
\begin{equation}
\boldsymbol{g}_{i}=\boldsymbol{p}_{i}-\boldsymbol{p}_{i+1}\,\,\left(1\le i\le N\right),\label{eq:72}
\end{equation}
which is apparently constant within the element. Hence, the \emph{Riemannian}
metric
\begin{equation}
g_{ij}=\boldsymbol{g}_{i}\cdot\boldsymbol{g}_{j}
\end{equation}
is also constant within the element. Moreover, when considering the
usual elastic bodies, $T_{\cdot k}^{i}$ is usually dependent on only
$g_{ij}$, then $T_{\cdot k}^{i}$ is also constant within the element.
As the result of above considerations, the integrated function is
constant within the element and the following formulations can be
used:
\begin{equation}
\boldsymbol{\omega}_{j}^{1}\left(T_{\cdot\gamma}^{\alpha}\right)\equiv\frac{1}{2}L_{j}\left[T_{\cdot1}^{1}g^{11}\nabla g_{11}\right]_{j},\label{eq:71-2}
\end{equation}
\begin{equation}
\boldsymbol{\omega}_{j}^{2}\left(T_{\cdot\gamma}^{\alpha}\right)\equiv\frac{1}{2}S_{j}\left[T_{\cdot\gamma}^{\alpha}g^{\gamma\beta}\nabla g_{\alpha\beta}\right]_{j}\,\,\,\,\left(1\le\alpha,\beta,\gamma\le2\right),
\end{equation}
\begin{equation}
\boldsymbol{\omega}_{j}^{3}\left(T_{\cdot\gamma}^{\alpha}\right)\equiv\frac{1}{2}V_{j}\left[T_{\cdot\gamma}^{\alpha}g^{\gamma\beta}\nabla g_{\alpha\beta}\right]_{j}\,\,\,\,\left(1\le\alpha,\beta,\gamma\le3\right),
\end{equation}
where $L_{j},S_{j},V_{j}$ respectively denote the length, the area,
and the volume of each dimensional element, namely they are given
by
\begin{equation}
L_{j}=\left.\sqrt{\det g_{11}}\right|_{j},
\end{equation}
\begin{equation}
S_{j}=\frac{1}{2}\left.\sqrt{\det g_{\alpha\beta}}\right|_{j}\,\,\left(1\le\alpha,\beta\le2\right),
\end{equation}
\textbf{
\begin{equation}
V_{j}=\frac{1}{6}\left.\sqrt{\det g_{\alpha\beta}}\right|_{j}\,\,\left(1\le\alpha,\beta\le3\right).
\end{equation}
}In addition, $g^{ij}$ is the inverse of $g_{ij}$. The inverses
of tiny matrices can be calculated by using the following explicit
representations:

\begin{equation}
\left(g_{11}\right)^{-1}=\frac{1}{g_{11}},
\end{equation}
\begin{equation}
\left[\begin{array}{cc}
g_{11} & g_{12}\\
g_{21} & g_{22}
\end{array}\right]^{-1}=\frac{1}{\det g_{ij}}\left[\begin{array}{cc}
g_{22} & -g_{12}\\
-g_{21} & g_{11}
\end{array}\right],
\end{equation}
{\footnotesize 
\begin{align}
\left[\begin{array}{ccc}
g_{11} & g_{12} & g_{13}\\
g_{21} & g_{22} & g_{23}\\
g_{31} & g_{32} & g_{33}
\end{array}\right]^{-1} & =\frac{1}{\det g_{ij}}\left[\left[\begin{array}{c}
g_{12}\\
g_{22}\\
g_{32}
\end{array}\right]\times\left[\begin{array}{c}
g_{13}\\
g_{23}\\
g_{33}
\end{array}\right]\right.\\
 & \left.\begin{array}{cc}
\left[\begin{array}{c}
g_{13}\\
g_{23}\\
g_{33}
\end{array}\right]\times\left[\begin{array}{c}
g_{11}\\
g_{21}\\
g_{31}
\end{array}\right] & \left[\begin{array}{c}
g_{11}\\
g_{21}\\
g_{31}
\end{array}\right]\times\left[\begin{array}{c}
g_{12}\\
g_{22}\\
g_{32}
\end{array}\right]\end{array}\right],
\end{align}
}where
\begin{equation}
\left[\begin{array}{c}
a_{1}\\
a_{2}\\
a_{3}
\end{array}\right]\times\left[\begin{array}{c}
b_{1}\\
b_{2}\\
b_{3}
\end{array}\right]\equiv\left[\begin{array}{c}
a_{2}b_{3}-b_{2}a_{3}\\
a_{3}b_{1}-b_{3}a_{1}\\
a_{1}b_{2}-b_{1}a_{2}
\end{array}\right].
\end{equation}
Thus, the tiny inverses have been completely eliminated from $\boldsymbol{\omega}$.

\subsection{Gradient vectors and the general form}

\textbf{}
\begin{figure}[!tbh]
\centering{}\textbf{\includegraphics{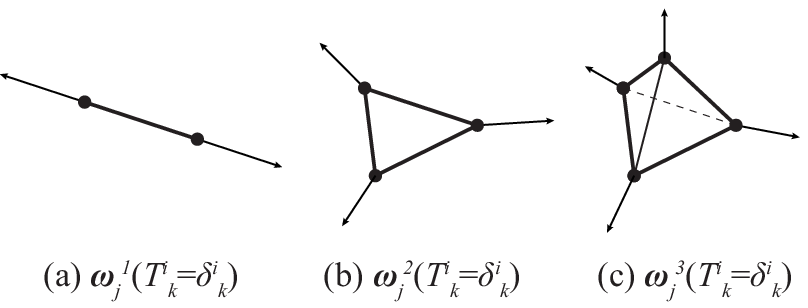}\caption{\label{fig:grad}$\boldsymbol{\omega}_{j}^{N}\left(T_{\cdot k}^{i}=\delta_{\cdot k}^{i}\right)$}
}
\end{figure}

In this subsection, the relation between $\boldsymbol{\omega}_{j}^{N}\left(T_{\cdot\gamma}^{\alpha}\right)$
and the gradient vectors are discussed. 

Interestingly, as are depicted in Fig. \ref{fig:grad}, when $T_{\cdot\gamma}^{\alpha}=\delta_{\cdot\gamma}^{\alpha}$,
the following exact relations are formed: 
\begin{equation}
\boldsymbol{\omega}_{j}^{1}\left(\delta_{\cdot\gamma}^{\alpha}\right)=\nabla L_{j},
\end{equation}
\begin{equation}
\boldsymbol{\omega}_{j}^{2}\left(\delta_{\cdot\gamma}^{\alpha}\right)=\nabla S_{j},
\end{equation}
\begin{equation}
\boldsymbol{\omega}_{j}^{3}\left(\delta_{\cdot\gamma}^{\alpha}\right)=\nabla V_{j}.
\end{equation}
The demonstrations of above relations can be obtained by altering
$\delta$ symbols into $\nabla$ symbols in the demonstration of that
the \textbf{minimal volume problem} is a special case of the \textbf{principle
of virtual work} such that $T_{\cdot\gamma}^{\alpha}=\delta_{\cdot\gamma}^{\alpha}$,
which was described in the subsection 3.2. Therefore, a set of $\left\{ \nabla L_{j},\nabla S_{j},\nabla V_{j}\right\} $
coincides with $\left\{ \boldsymbol{\omega}_{j}^{1}\left(T_{\cdot\gamma}^{\alpha}\right),\boldsymbol{\omega}_{j}^{2}\left(T_{\cdot\gamma}^{\alpha}\right),\boldsymbol{\omega}_{j}^{3}\left(T_{\cdot\gamma}^{\alpha}\right)\right\} $
when $T_{\cdot\gamma}^{\alpha}=\delta_{\cdot\gamma}^{\alpha}$, and
then $\boldsymbol{\omega}_{j}^{N}\left(T_{\cdot\gamma}^{\alpha}\right)$
is one of the natural generalizations of $\left\{ \nabla L_{j},\nabla S_{j},\nabla V_{j}\right\} $.
Furthermore, $\boldsymbol{\omega}_{j}^{N}\left(T_{\cdot\gamma}^{\alpha}\right)$
can be used when $\left\{ \nabla L_{j},\nabla S_{j},\nabla V_{j}\right\} $
are calculated.

Based on above considerations, it can be noticed that only special
cases such that $T_{\cdot\gamma}^{\alpha}$ is given as just a scalar
multiple of $\delta_{\cdot\gamma}^{\alpha}$ have been discussed in
the section 2. Therefore, it is very natural to consider general functions
as $T_{\cdot\gamma}^{\alpha}$. Particularly, a map from $g_{ij}$
to $T_{\cdot\gamma}^{\alpha}$ is no other than the constitutive law
it self.

The explicit representations of $\nabla L_{j}$ and $\nabla S_{j}$
are presented in Appendix A and one may notice that they look very
different while the difference between $\boldsymbol{\omega}_{j}^{1}\,\mathrm{and}\,\boldsymbol{\omega}_{j}^{2}$
is just the dimensions of the matrices; however, due to $g^{ij}$
which is defined as inverse matrix, the apparent difference between
$\nabla L_{j}$ and $\nabla S_{j}$ is resulted from the difference
between the explicit representations of tiny inverse matrices.

By the way, each $\boldsymbol{\omega}_{j}^{N}$ is just a mixture
of the gradient vectors $\nabla g_{ij}$, then even though a function
$f_{j}$ such that $\boldsymbol{\omega}_{j}^{N}=\nabla f_{j}$, is
not found in usual, $\boldsymbol{\omega}_{j}^{N}$ is a row vector
that highly resemble the gradient vectors. Therefore, the \textbf{discrete
stationary condition} which was formulated in the subsection 3.3 is
expected to be solved by the \textbf{two-term} or the \textbf{three-term
method} by just altering $\nabla\Pi$ into $\boldsymbol{\omega}$.

\subsection{Numerical examples}

In this subsection, some examples that discrete \textbf{stationary
conditions} can be solved by the \textbf{three-term method} are illustrated.
As the simplest constitutive law, only

\begin{eqnarray}
T_{\cdot k}^{i} & = & Eg^{il}\left(g_{lk}-\bar{g}_{lk}\right),\\
\therefore T^{ij}=T_{\cdot k}^{i}g^{kj} & = & Eg^{il}\left(g_{lk}-\bar{g}_{lk}\right)g^{kj},
\end{eqnarray}
is considered, which is the uniform linear material with Poisson ratio=0,
and where $E$ is the stiffness factor. It must be remarked that the
stiffness factor $E$ is identical with Young's modulus only when
$N=3$, otherwise it is multiplied with the sectional area or the
thickness, when a \emph{Riemannian} manifold is related with a real
material. Additionally, $e_{ik}=\left(g_{ik}-\bar{g}_{ik}\right)$
is no other than the strain tensor itself and $\bar{g}_{lk}$ is the
\emph{Riemannian} metric treated as constant and is measured on the
initial shape on which the stress tensor vanishes. Note that while
$T_{\cdot k}^{i}$ is not a symmetric matrix, $T^{ij}$ is a symmetric
matrix.

Unlike the numerical examples described in the section 2, in each
initial step of the following numerical examples, $\left\{ x_{1},\cdots,x_{n}\right\} $
were not given by random numbers but were given by the coordinates
of the initial shape, and $\bar{g}_{ik}$ were calculated on such
initial shapes.

\begin{figure}[!tbh]
\centering{}\includegraphics{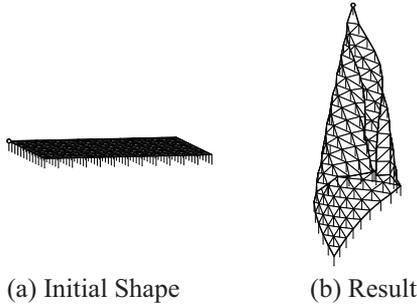}\caption{\label{fig:Natural-Forms-of}Natural Forms of Handkerchief 1}
\end{figure}

\begin{figure}[!tbh]
\centering{}\includegraphics{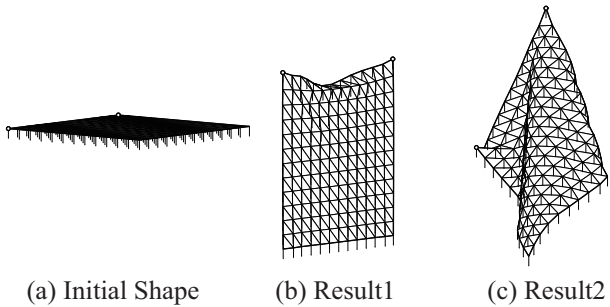}\caption{\label{fig:Natural-Forms-of-1}Natural Forms of Handkerchief 2}
\end{figure}

Fig. \ref{fig:Natural-Forms-of} and \ref{fig:Natural-Forms-of-1}
show natural forms of handkerchief that are hanged by 1, or 2 point
under gravity. The dimension of the numerical model is 8.0x8.0, and
every z-components of the nodal forces were set as 0.1 . Each form
has been obtained by solving

\begin{equation}
\boldsymbol{\omega}=\sum_{j}\boldsymbol{\omega}_{j}^{2}\left(T_{\cdot k}^{i}\right)-\boldsymbol{p}=\boldsymbol{0},
\end{equation}
which is a \textbf{discrete stationary condition} or a discrete form
of equilibrium equation, by the \textbf{three-term method}. In addition,
$E=50$ as the stiffness factor and $\alpha=0.2$ as the step-size
factor were used.

\begin{figure}[!tbh]
\centering{}\includegraphics{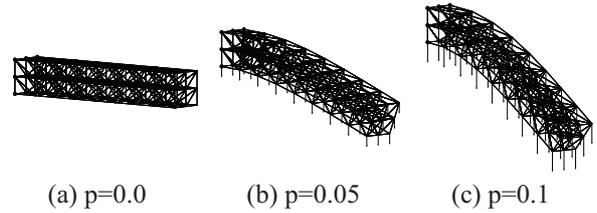}\caption{\label{fig:Large-deformations-of}Large deformations of a cantilever
under gravity}
\end{figure}

\begin{figure}[!tbh]
\centering{}\includegraphics{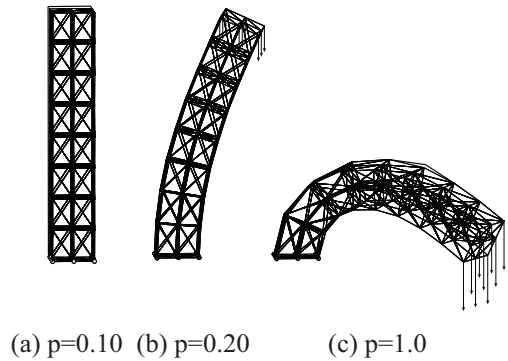}\caption{\label{fig:Large-deformation-after}Large deformation after buckling}
\end{figure}

Fig. \ref{fig:Large-deformations-of} shows the large deformations
of a cantilever under gravity whose dimension is 2.0-2.0-12.0. Fig.
\ref{fig:Large-deformation-after} shows the large deformations after
buckling of a bar which has the same dimension of the former . Each
form has been obtained by solving

\begin{equation}
\boldsymbol{\omega}=\sum_{j}\boldsymbol{\omega}_{j}^{3}\left(T_{\cdot k}^{i}\right)-\boldsymbol{p}=\boldsymbol{0},
\end{equation}
 which is the \textbf{discrete stationary condition}, by the \textbf{three-term
method}. In both analyses, $E=50$ as the stiffness factor and $\alpha=0.2$
as the step-size factor were used.

In the analysis which resulted Fig. \ref{fig:Large-deformations-of},
every z-components of the nodal forces were set as $p$, which is
shown in the figure. In the analysis which resulted Fig. \ref{fig:Large-deformation-after},
small random numbers were firstly supplemented to the initial nodal
coordinates to make the model easily buckle. Then, z-components of
the nodal forces of only 9 nodes which place on the top of the model
were set as $p$, which is shown in the figure. Even if this can be
explained as one kind of buckling phenomena, the analysis itself is
just a large deformation analysis; hence precise identify of critical
load is almost impossible. However, the Euler buckling load corresponding
to this example was calculated as $p_{cr}=1.14$ and its division
by 9 is $0.126$, which indeed places between Fig. \ref{fig:Large-deformation-after}
(a) and (b).

\section{Conclusions}

In the first half of this work, the direct minimization approaches
were discussed, in which some form finding problems of tension structures
were considered. Especially, as the standard strategies of for direct
minimization approaches, the \textbf{two-term method}, the \textbf{three-term
method}, and the \textbf{dual estimate} were presented. In addition,
the relation over the \textbf{principle of virtual work}, the \textbf{stationary
condition}, and the \textbf{standard search direction} were clarified,
which are the means of direct minimization approaches.

In the last half of this work, starting from the \textbf{principle
of virtual work }(field equation) that usually appeared in the continuum
mechanics, the discrete \textbf{principle of virtual work} was deduced.
Moreover, the the discrete \textbf{stationary condition} and the \textbf{standard
search direction} were formulated to let the \textbf{three-term method}
feasible. Those formulae were expressed with $\boldsymbol{\omega}_{j}^{N}\left(T_{\cdot k}^{i}\right)$,
which is one of the natural generalizations of $\left\{ \nabla L_{j},\nabla S_{j},\nabla V_{j}\right\} $,
hence, the last half of this work was a generalization of the first
half of this work. Finally, some large deformation analyses of continuum
bodies were illustrated. Those various types of numerical examples
which were shown in this work imply the potential ability of the \textbf{three-term
method} that can be a powerful means of solving various types of static
problems of continuum bodies.

\section*{Acknowledgments}

This research was partially supported by the Ministry of Education,
Culture, Sports, Science and Technology, Grant-in-Aid for JSPS Fellows,
10J09407, 2011

\appendix

\section{Gradients}

\subsection{Gradient of Linear Element Length}

\label{sub:grad_length}

Suppose p and q denote two nodes. Let

\noindent 
\begin{equation}
\boldsymbol{p}\equiv\left[\begin{array}{c}
p_{x}\\
p_{y}\\
p_{z}
\end{array}\right]\,\mathrm{and}\,\boldsymbol{q}\equiv\left[\begin{array}{c}
q_{x}\\
q_{y}\\
q_{z}
\end{array}\right]
\end{equation}
represent the Cartesian coordinates of p and q.

The length of the line determined by p and q is given by 
\begin{align}
 & L\left(p_{x},\, p_{y},\, p_{z},\, q_{x},\, q_{y},\, q_{z}\right)\label{eq:length_func}\\
 & \equiv\sqrt{{(p_{x}-q_{x})^{2}+(p_{y}-q_{y})^{2}+(p_{z}-q_{z})^{2}}}.
\end{align}

\begin{figure}[!tbh]
\centering{}\textbf{\includegraphics{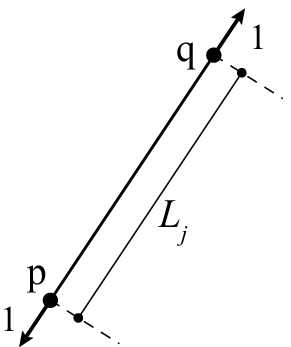}\caption{\label{fig:L}$\hat{\nabla}L_{j}$}
}
\end{figure}

If the gradient of $L$ is defined by{\scriptsize 
\begin{align}
\hat{\nabla}L & \equiv\left[\frac{\partial L}{\partial p_{x}},\frac{\partial L}{\partial p_{y}},\frac{\partial L}{\partial p_{z}},\frac{\partial L}{\partial q_{x}},\frac{\partial L}{\partial q_{y}},\frac{\partial L}{\partial q_{z}}\right],
\end{align}
}{\scriptsize \par}

\noindent its components are as follows:{\scriptsize 
\begin{align}
\hat{\nabla}L & =\left[\frac{p_{x}-q_{x}}{L},\frac{p_{y}-q_{y}}{L},\frac{p_{z}-q_{z}}{L},\frac{q_{x}-p_{x}}{L},\frac{q_{y}-p_{y}}{L},\frac{q_{z}-p_{z}}{L}\right],\label{eq:gradient_L}
\end{align}
}and its visualization is presented by Fig. \ref{fig:L}.

\begin{figure}[!tbh]
\centering{}\includegraphics{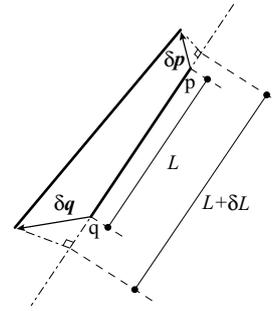}\caption{\label{figLinear element2}Variation of Element Length}
\end{figure}

Let us investigate $\delta L$, i.e.
\begin{equation}
\delta L\equiv\hat{\nabla}L\cdot\left[\begin{array}{c}
\delta\boldsymbol{p}\\
\delta\boldsymbol{q}
\end{array}\right].
\end{equation}
As shown in Fig. \ref{figLinear element2}, $\delta\boldsymbol{p}$
and $\delta\boldsymbol{q}$ are firstly projected to the line determined
by p and q, then, $\delta L$ is measured on the line. 

\textbf{}

\subsection{Gradient of Triangular Element Area}

Let p, q, and r be three vertices. Let 

\textbf{
\begin{equation}
\boldsymbol{p}\equiv\left[\begin{array}{c}
p_{x}\\
p_{y}\\
p_{z}
\end{array}\right],\,\boldsymbol{q}\equiv\left[\begin{array}{c}
q_{x}\\
q_{y}\\
q_{z}
\end{array}\right],\,\boldsymbol{r}\equiv\left[\begin{array}{c}
r_{x}\\
r_{y}\\
r_{z}
\end{array}\right],
\end{equation}
}denote the Cartesian coordinates of p, q, and r.

The area of the triangle determined by p, q, and r is given by 
\begin{align}
S(p_{x},\cdots,r_{z})\equiv & \frac{1}{2}\sqrt{\boldsymbol{N}\cdot\boldsymbol{N}},\\
\left(\boldsymbol{N}\right.\equiv & \left.\left(\boldsymbol{q}-\boldsymbol{p}\right)\times\left(\boldsymbol{r}-\boldsymbol{p}\right)\right).
\end{align}

\begin{figure}[!tbh]
\centering{}\textbf{\includegraphics{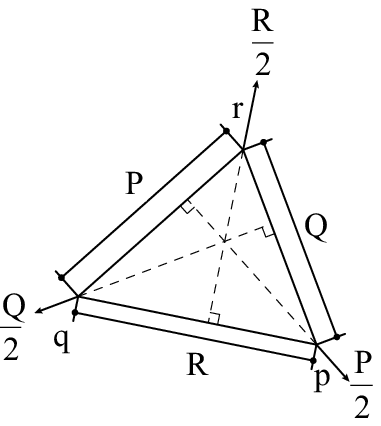}\caption{\textbf{\label{fig:S}$\hat{\nabla}S_{j}$}}
}
\end{figure}

If the gradient of $S$ is defined by

\begin{equation}
\hat{\nabla}S\equiv\left[\begin{array}{ccccc}
\frac{\partial S}{\partial p_{x}}, & \frac{\partial S}{\partial p_{y}}, & \frac{\partial S}{\partial p_{z}}, & \cdots & ,\frac{\partial S}{\partial r_{z}}\end{array}\right],
\end{equation}
its components are as follows:{\scriptsize 
\begin{alignat}{1}
\hat{\nabla}S=\frac{1}{2}\boldsymbol{n}\cdot & \left[(\boldsymbol{r}-\boldsymbol{q})\times\left\{ \left[\begin{array}{c}
1\\
0\\
0
\end{array}\right],\left[\begin{array}{c}
0\\
1\\
0
\end{array}\right],\left[\begin{array}{c}
0\\
0\\
1
\end{array}\right]\right\} \right.\nonumber \\
 & \left.,(\boldsymbol{p}-\boldsymbol{r})\times\left\{ \left[\begin{array}{c}
1\\
0\\
0
\end{array}\right],\left[\begin{array}{c}
0\\
1\\
0
\end{array}\right],\left[\begin{array}{c}
0\\
0\\
1
\end{array}\right]\right\} \right.\label{eq:grad_S}\\
 & \left.,(\boldsymbol{q}-\boldsymbol{p})\times\left\{ \left[\begin{array}{c}
1\\
0\\
0
\end{array}\right],\left[\begin{array}{c}
0\\
1\\
0
\end{array}\right],\left[\begin{array}{c}
0\\
0\\
1
\end{array}\right]\right\} \right],\nonumber 
\end{alignat}
}where $\boldsymbol{n}$ is defined by

\begin{equation}
\boldsymbol{n}\equiv\frac{\boldsymbol{N}}{\left|\boldsymbol{N}\right|},
\end{equation}
and a visualization of $\hat{\nabla}S$ is presented by Fig. \ref{fig:S}.

\begin{figure}[!tbh]
\centering{}\includegraphics{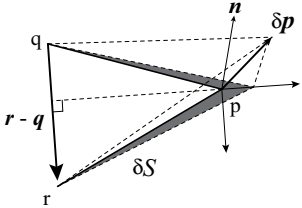}\caption{\label{fig:Triangular-Element2}Variation of Element Area}
\end{figure}

Let us investigate $\delta S$, i.e.{\scriptsize 
\begin{align}
\delta S & =\frac{1}{2}\boldsymbol{n}\cdot\left((\boldsymbol{r-q})\times\delta\boldsymbol{p}+(\boldsymbol{p}-\boldsymbol{r})\times\delta\boldsymbol{q}+(\boldsymbol{q}-\boldsymbol{p})\times\delta\boldsymbol{r}\right).\label{eq:deltaS}
\end{align}
}With respect to $\delta\boldsymbol{p}$, for example, when $\delta\boldsymbol{p}$
is orthogonal to the element, $\left(\boldsymbol{r}-\boldsymbol{q}\right)\times\delta\boldsymbol{p}$
becomes orthogonal to $\boldsymbol{n}$, then $\delta S$ vanishes
(see Fig. \ref{fig:Triangular-Element2}). On the other hand, when
$\delta\boldsymbol{p}$ is parallel to the opposite side, $\left(\boldsymbol{r}-\boldsymbol{q}\right)\times\delta\boldsymbol{p}$
vanishes, then $\delta S$ vanishes. Therefore, only the component
of $\delta\boldsymbol{p}$ which is parallel to the perpendicular
line from p to the opposite side can produce $\delta S$. In other
words, $\delta S$ is measured on the plane determined by p, q, and
r.

\textbf{}

\subsection{Gradient of Riemannian Metrics}

The explicit representation of $\nabla g_{ij}$ can be obtained by
referring the following calculations. First, Eq. \eqref{eq:71-1}
and Eq. \eqref{eq:72} follows{\scriptsize 
\begin{equation}
dg_{ij}=d\left(\boldsymbol{p}_{i+1}-\boldsymbol{p}_{i}\right)\cdot\left(\boldsymbol{p}_{j+1}-\boldsymbol{p}_{j}\right)+\left(\boldsymbol{p}_{i+1}-\boldsymbol{p}_{i}\right)\cdot d\left(\boldsymbol{p}_{j+1}-\boldsymbol{p}_{j}\right),\,\,\left(1\le i,j\le N\right)
\end{equation}
}which can be expanded as{\scriptsize 
\begin{align}
 & dg_{ij}=\nonumber \\
 & dX_{i+1}\left(X_{j+1}-X_{j}\right)-dX_{i}\left(X_{j+1}-X_{j}\right)+dX_{j+1}\left(X_{i+1}-X_{i}\right)-dX_{j}\left(X_{i+1}-X_{i}\right)\nonumber \\
 & +dY_{i+1}\left(Y_{j+1}-Y_{j}\right)-dY_{i}\left(Y_{j+1}-Y_{j}\right)+dY_{j+1}\left(Y_{i+1}-Y_{i}\right)-dY_{j}\left(Y_{i+1}-Y_{i}\right)\nonumber \\
 & +dZ_{i+1}\left(Z_{j+1}-Z_{j}\right)-dZ_{i}\left(Z_{j+1}-Z_{j}\right)+dZ_{j+1}\left(Z_{i+1}-Z_{i}\right)-dZ_{j}\left(Z_{i+1}-Z_{i}\right),\label{eq:82}
\end{align}
}where $\left\{ X_{i},Y_{i},Z_{i}\right\} \,\left(1\le i\le N+1\right)$
represents the Cartesian coordinates of $i$-th node.

When the independent parameters $\left\{ x_{1},\cdots,x_{n}\right\} $
are selected as the Cartesian coordinates of all the free nodes $\left\{ X_{i},Y_{i},Z_{i}\right\} $
$\left(1\le i\le N+1\right)$, by comparing Eq. \eqref{eq:82} and
the following relation, the explicit representation of $\nabla g_{ij}$
can be obtained. 
\begin{equation}
dg_{ij}=\left[\frac{\partial g_{ij}}{\partial x_{1}}\cdots\frac{\partial g_{ij}}{\partial x_{n}}\right]\left[\begin{array}{c}
dx_{1}\\
\vdots\\
dx_{n}
\end{array}\right]=\nabla g_{ij}\cdot\left[\begin{array}{c}
dx_{1}\\
\vdots\\
dx_{n}
\end{array}\right].
\end{equation}

\end{document}